\documentclass[aps,twocolumn,superscriptaddress,floatfix,nofootinbib,prb]{revtex4-1}

\usepackage{lipsum}
\usepackage{graphicx}
\usepackage{amsmath,amssymb}
\usepackage{braket}
\usepackage{mathtools}
\usepackage{hyperref}

\usepackage{color}
\usepackage[normalem]{ulem}
\usepackage{amsfonts}
\newcommand{\veff}{v^{\text{eff}}}
\newcommand{\rhop}{\rho}
\newcommand{\rhot}{\rho^T}
\newcommand{\latt}{a}

\newcommand{\rap}{\lambda}
\newcommand{\vavg}{v^{\star}}

\def\Ket#1{\left|#1\right>}

\def\beq{\begin{equation}}
\def\eeq{\end{equation}}
\def\bea{\begin{eqnarray}}
\def\eea{\end{eqnarray}}

\begin{document}
\title{Generalized hydrodynamics, quasiparticle diffusion, and anomalous local relaxation in random integrable spin chains}
\author{Utkarsh Agrawal}
\affiliation{Department of Physics, University of Massachusetts, Amherst, MA 01003, USA}
\author{Sarang Gopalakrishnan}
\affiliation{Department of Physics and Astronomy, CUNY College of Staten Island, Staten Island, NY 10314;  Physics Program and Initiative for the Theoretical Sciences, The Graduate Center, CUNY, New York, NY 10016, USA}
\author{Romain Vasseur}
\affiliation{Department of Physics, University of Massachusetts, Amherst, MA 01003, USA}

\begin{abstract}
We study the nonequilibrium dynamics of random spin chains that remain integrable (i.e., solvable via Bethe ansatz): because of correlations in the disorder, these systems escape localization and feature ballistically spreading quasiparticles. We derive a generalized hydrodynamic theory for dynamics in such random integrable systems, including diffusive corrections due to disorder, and use it to study non-equilibrium energy and spin transport. We show that diffusive corrections to the ballistic propagation of quasiparticles can arise even in noninteracting settings, in sharp contrast with clean integrable systems. This implies that operator fronts broaden diffusively in random integrable systems. By tuning parameters in the disorder distribution, one can drive this model through an unusual phase transition, between a phase where all wavefunctions are delocalized and a phase in which low-energy wavefunctions are quasi-localized (in a sense we specify). Both phases have ballistic transport; however, in the quasi-localized phase, local autocorrelation functions decay with an anomalous power law, and the density of states diverges at low energy.
\end{abstract}
\maketitle
\vspace{1cm}

\section{Introduction}

The study of the dynamics of isolated, many-body quantum systems far from thermal equilibrium  has attracted a lot of attention recently, fueled by recent experimental developments on ultracold atoms~\cite{bloch_many-body_2008,kinoshita_quantum_2006,langen_experimental_2015}, trapped ions~\cite{leibfried2003single-ions, duan2010network-ions, Blatt2012SimulationsIons}, nitrogen-vacancy centers~\cite{Doherty2013NVreview, Schirhagl2014NVreview} and superconducting qubits~\cite{kelly2015state} platforms. Addressing questions about non-equilibrium transport, thermalization and far-from-equilibrium dynamics pose notable challenges for theory as they are not susceptible to the general principles and methods that govern the physics of low-energy, equilibrium systems. 

With the notable exception of many-body localized systems~\cite{ARCMP,VasseurMoore2016MBLReview,AbaninRev}, generic many-body systems are expected to be ``chaotic'', and to thermalize under their own dynamics~\cite{Polkovnikov-rev}. This process can be understood as the scrambling of quantum information as it becomes non-local and inaccessible to physical, local measurements. After a local equilibration regime, thermalizing systems can be well-described by classical hydrodynamic equations associated with conserved quantities -- typically, energy, particle number and momentum. These hydrodynamic equations describe the evolution of the system from local to global equilibrium. Another class of systems that escape thermalization in the traditional sense are quantum integrable systems,  including experimentally relevant examples like the Heisenberg antiferromagnet and the Lieb-Liniger Bose gas in one dimension~\cite{kinoshita_quantum_2006,langen_experimental_2015,PhysRevX.8.021030}. Such systems have stable quasiparticle excitations even at high temperature, and they possess an extensive number of conserved quantities which strongly constrain their dynamics, and prevent them from thermalizing like generic chaotic systems~
\cite{Calabrese:2006, PhysRevLett.106.217206, PhysRevLett.110.257203, 1742-5468-2013-07-P07012, Essler_2016, PhysRevLett.111.057203, PhysRevLett.113.117202, PhysRevLett.115.120601, PhysRevLett.115.157201, 2016arXiv160300440I,PhysRevB.89.125101}.
However, contrary to many-body localized systems, integrable systems do thermalize in a generalized sense, as they eventually reach 
a maximum entropy steady state described by a generalized Gibbs ensemble (GGE)~\cite{rigol_relaxation_2007,caux_constructing_2012,rigol_breakdown_2009,langen_experimental_2015}. Such steady-states can exhibit non-zero currents, and are commonly referred to as non-equilibrium steady states (NESS) in the literature~\cite{Bernard_2016, VasseurMoore2016MBLReview}, even though they are natural equilibrium states for integrable systems. 

A major step in understanding the non-equilibrium dynamics of quantum integrable systems  was the formulation of what is now known as ``generalized hydrodynamics''~(GHD)~\cite{castro-alvaredo_emergent_2016,bertini_transport_2016}, which are Euler hydrodynamics equations (0th order hydrodynamics) obtained in the large space-time limit where the system is locally in equilibrium. While the prospect of solving infinitely-many hydrodynamic equations (one for each conserved quantity in the system) originally appeared daunting, GHD can be conveniently formulated in the basis of quasiparticle excitations: in that language, they can be naturally interpreted as describing a semi-classical gas of solitons (quasiparticles)~\cite{EL2003374,PhysRevLett.95.204101,BBH,solitongases}. The key ingredient of GHD is the effective group velocity $v^{\rm eff}$ of the quasiparticles~\cite{PhysRevLett.113.187203,castro-alvaredo_emergent_2016,bertini_transport_2016} which depends on the density of all the other quasiparticles in the presence of interactions: at the semi-classical level, quasiparticle wavepackets  pick up a phase shift when they collide, leading to a Wigner time-delay. This approach was successfully applied to two-reservoir setups~\cite{castro-alvaredo_emergent_2016,bertini_transport_2016} and more generally to locally equilibrated inhomogeneous initial states~\cite{bulchandani_solvable_2017, doyon_large-scale_2017,piroli_transport_2017}, and has helped addressing a number of key questions in the field concerning Drude weights~\cite{PhysRevLett.119.020602, BBH,GHDII,PhysRevB.96.081118,PhysRevB.97.081111}, external potentials and traps~\cite{SciPostPhys.2.2.014,PhysRevLett.120.164101,2017arXiv171100873C}, correlation functions~\cite{10.21468/SciPostPhys.5.5.054}, entanglement dynamics~\cite{alba2017entanglement}, or even large-deviation functions~\cite{2018arXiv181202082M}. 

The GHD framework was recently generalized to include diffusive effects in interacting integrable models~\cite{de_nardis_diffusion_2018, gopalakrishnan_hydrodynamics_2018,de_nardis_hydrodynamic_2018, 2018arXiv181202701G} -- corresponding to ``1st order'' or Navier-Stokes hydrodynamics, with important consequences for the nature of spin transport in XXZ spin chains~\cite{de_nardis_hydrodynamic_2018, 2018arXiv181202701G}. In particular, Ref.~\onlinecite{de_nardis_diffusion_2018} provided a general exact expression of the ``diffusion matrix'' of the quasiparticles using a form factor expansion of the Kubo formula. Intuitively, diffusive corrections can be seen to arise as follows~\cite{gopalakrishnan_hydrodynamics_2018}: the effective velocity of a given quasiparticle depends on the density of quasiparticles in a mean-field fashion that ignores fluctuations; reintroducing thermal fluctuations naturally leads to a diffusive broadening of quasiparticle trajectories in a generic GGE state. Two key ingredients are needed for such diffusive corrections to be present: density-dependent velocities and thermal fluctuations (non-zero entropy states). This immediately implies that in non-interacting integrable models where the group velocity is obtained from band theory and is independent of density, there should be no diffusion~\cite{spohn_interacting_2018}. For non-interacting systems, it is thus natural to expect that the lowest order correction to ballistic GHD comes from higher order derivative terms, which lead to $t^{1/3}$ spreading governed by the Airy kernel (see Ref.~\onlinecite{fagotti_higher-order_2017} and references therein). Clearly, such higher-order corrections are subleading in the presence of diffusive $t^{1/2}$ spreading. 

In this paper, we study a class of integrable random spin chains which support diffusive corrections even in the absence of interactions. 
These spin chains are a special limit of a more general class of random interacting spin chains that remain integrable. In one-dimensional free-particle problems, disorder generically leads to Anderson localization. Though Anderson localized systems are ``integrable'' in a sense, here we will use the term ``integrability'' exclusively to refer to Bethe-ansatz solvable systems with stable ballistically propagating quasiparticles. There are examples of integrable models with impurities~\cite{vega_new_1992,de_vega_thermodynamics_1994,essler_integrable_2018} where disorder is correlated in such a way that integrability in this sense is preserved. Such systems were recently shown to exhibit ballistic transport even at strong disorder in Ref.~\onlinecite{essler_integrable_2018}. It is natural to expect such random systems to exhibit diffusive corrections to ballistic transport even without interactions, as quasiparticles scatter off random static impurities and thus undergo biased random walks. These models illustrate that whereas non-interacting systems are often said to be always integrable, integrability for a random system leads to correlated disorder that can allow one-dimensional random systems to escape Anderson localization. Although most states in these models are only weakly affected by the disorder, quasiparticle states near energy $|E| = 0$ have properties that are sensitive to the tails of the disorder distribution. We find disorder distributions for which these quasiparticles have vanishing velocities, so that the behavior of local autocorrelation functions is anomalous. We find that the onset of anomalous behavior in the disorder-averaged autocorrelation functions is associated with the onset of a divergence in the density of states at zero energy, as well as a form of quasi-localization of the low-energy wavefunctions that we discuss below. We compute this anomalous relaxation exponent using GHD. There are strong local correlations between the density of states and the quasiparticle velocity, as both are dominated by rare regions; our work shows how GHD can be adapted to incorporate these rare region effects.

In this work, we propose a hydrodynamic theory to describe such random integrable spin chains, including diffusive corrections due to disorder.  The plan of this paper is as follows: in Section~\ref{SectionModel}, we recall the definition of a family of random integrable spin chains recently studied in Ref.~\onlinecite{essler_integrable_2018}, and we briefly review their thermodynamics. 
We then formulate a coarse-grained GHD theory for the dynamics of such systems in Section~\ref{GHD}, with an emphasis on diffusive corrections due to disorder in non-interacting settings (for which disorder is the only possible source of diffusive corrections). This framework is applied to study non-equilibrium spin and energy transport, and the predictions are compared to numerical results in Section~\ref{sec:Transport}. 
Transport is dominated by fast quasiparticles with energies well away from $|E| = 0$. We turn in Sec.~\ref{SecWavefunctions} to a more careful discussion of states near $|E| = 0$. For these states we find a quasi-localization transition; in the quasi-localized regime, low-energy wavefunctions consist of a few local peaks, quasiparticle velocities vanish, and the density of states diverges in the $|E| \rightarrow 0$ limit.
Consequences for operator spreading and scrambling are briefly discussed in Section~\ref{sec:OTOC}, and a discussion and outlooks for future works are gathered in Section~\ref{sec:Discussion}.

\section{Random integrable spin chains}

In this section, we introduce a family of random integrable spin chains, following closely Ref.~\onlinecite{essler_integrable_2018}, and we briefly review their thermodynamic Bethe ansatz solution. 

\label{SectionModel}

\subsection{Hamiltonian}

Let us consider a random XXZ spin-$\frac{1}{2}$ chain $H = \sum_i J_i \left[ \vec{\sigma}_{i}.\vec{\sigma}_{i+1} \right]_{\Delta_i}$, where $\left[ \vec{\sigma}_{j}.\vec{\sigma}_{k} \right]_{\Delta_i}$ is a shorthand notation for $\sigma^x_j\sigma^x_k+\sigma^y_j\sigma^y_k+\Delta_i(\sigma^z_j\sigma^z_k-1)$. In the clean (homogeneous) case, this model is integrable, but the introduction of disorder immediately breaks integrability, and leads to a model that is either chaotic or many-body localized~\cite{PhysRevLett.110.067204,PhysRevB.93.134207}. However, it is possible to preserve integrability~\cite{vega_new_1992,de_vega_thermodynamics_1994,essler_integrable_2018} by introducing next-to-nearest neighbor interactions, and by carefully choosing the inhomogeneous couplings:
\begin{align}
H=\sum_{j=1}^{L/2} & J^{(1)}_{2j}\left([\vec{\sigma}_{2j-1}.\vec{\sigma}_{2j}]_{\Delta_{2j}} + [\vec{\sigma}_{2j}.\vec{\sigma}_{2j+1}]_{\Delta_{2j}}  \right) \label{eq:Hamil in spin} \nonumber \\
                   & +K_{2j}\left( [\vec{\sigma}_{2j}.\left(\vec{\sigma}_{2j-1}\times \vec{\sigma}_{2j+1} \right)]_{\Delta_{2j}^{-1}}+\Delta_{2j}^{-1}\right) \nonumber \\
                   & +J^{(2)}_{2j}\left(\vec{\sigma}_{2j-1}.\vec{\sigma}_{2j+1}-1\right).
\end{align}
The first line of the hamiltonian corresponds to an XXZ interaction, while the last line is an isotropic Heinsenberg interaction. The middle line is more unusual, as it involves three spins. The parameters in the hamiltonian are given by
\begin{align}
J^{(1)}_{2j}&=\frac{\rm{sin}^2\eta\ \rm{cosh}\xi_{2j}}{\rm{sin}^2\eta+\rm{sinh}^2\xi_{2j}},\ \ J^{(2)}_{2j}=\frac{\rm{cos}\ \eta\ \rm{sinh}^2\xi_{2j}}{\rm{sin}^2\eta+\rm{sinh}^2\xi_{2j}}, \nonumber \\
K_{2j}&=\frac{\rm{sin}\ \eta\ \rm{cos}\ \eta\ \rm{sinh}\xi_{2j}}{\rm{sin}^2\eta+\rm{sinh}^2\xi_{2j}},\ \ \Delta_{2j}=\frac{\rm{cos}\ \eta}{\rm{cosh}\ \xi_{2j}},  \label{eq:Random parameters}
\end{align}
with $\xi_{2j}$ a random coupling, while $\eta$ is an overall global parameter that parametrizes the interaction strength. For $\xi_{2j}=0$, one recovers the usual XXZ spin chain. A remarkable feature of this model is that it remains integrable for any choice of the inhomogeneous couplings $\xi_{2j}$. 
Away from the zero-energy limit, the properties of this model are insensitive to details of the disorder distribution.
Therefore, except as specified below (i.e., in Sec.~\ref{SecWavefunctions} and subsequently), we will take the $\xi_{2j}$ couplings to be random variables drawn from the gaussian distribution
\begin{equation}
P(\xi)=\frac{1}{\sqrt{2\pi W^2}}e^{-\xi^2/2W^2}.
\end{equation} 
Later, we will also consider the exponential distribution
\beq\label{expdis}
P(\xi) = \frac{\phi}{2} e^{-\phi |\xi|}.
\eeq
for which a sharp quasi-localization transition exists.

We will be especially interested in the special point $\eta=\pi/2$. For this value of $\eta$, the XXX part of the Hamiltonian is set to zero, leaving behind a random XX model with three spin interactions,
\begin{align}
H=&\sum_{j=1}^{L/2}\Bigg[   \frac{1}{\rm{cosh}\ \xi_{2j}} \sum_{\alpha=x,y} \left[\sigma^\alpha_{2j-1}\sigma^\alpha_{2j}+\sigma^\alpha_{2j}\sigma^\alpha_{2j+1}\right] \label{eq:Free ham in spin}\\          
		      &	+  \rm{tanh}(\xi_{2j})\left[\sigma^y_{2j-1}\sigma^z_{2j}\sigma^x_{2j+1}-\sigma^x_{2j-1}\sigma^z_{2j}\sigma^y_{2j+1}\right]         \Bigg].  \nonumber
\end{align}
The above hamiltonian can be diagonalized via Jordan-Wigner transformation reducing it to a free fermion model
\begin{align}
H=&-\sum_{j=1}^{L}  \frac{2}{\rm{cosh}(\xi_{j})} \left( c^\dagger_{j}c_{j+1}+{\rm h.c.}\right)  \notag \label{eq:Hamiltonian}\\
  &+ \sum_{j=1}^{L/2}2 i \rm{tanh}(\xi_{2j})  \left( c^\dagger_{2j-1}c_{2j+1} - {\rm h.c.} \right),  
\end{align} 
where the $\xi_j$'s are random parameters used in \eqref{eq:Hamil in spin}, extended to odd sites via relation, $\xi_{2j-1}=\xi_{2j}$. We use periodic boundary conditions for the fermions for an even number of sites. While non-interacting and disordered, this model was shown to escape Anderson localization in Ref.~\onlinecite{essler_integrable_2018}, and to exhibit ballistic transport of conserved quantities. It is then natural to ask if transport properties of this model can be captured using generalized hydrodynamic equations, properly adapted to deal with the quenched disorder. If so, it is natural to expect disorder to lead to new hydrodynamic effects, such as diffusion. In the following, we will mostly focus on the special point $\eta = \pi/2$ (eqs.~\eqref{eq:Hamiltonian}~\eqref{eq:Free ham in spin}), though we expect our approach to generalize to any value of $\eta$. This will be convenient as the free fermion representation of this model allows one to simulate numerically the non-equilibrium dynamics of this system easily, and more importantly, the absence of interactions will allow us to isolate the effect of disorder on diffusion. 

\subsection{Thermodynamics\label{sec:TBA}}

The model introduced above admits an exact solution by Bethe Ansatz. In the following, we very briefly review the Thermodynamic Bethe Ansatz (TBA) approach to integrable systems, with the main goal of introducing some notation and language that will be used in the remainder of this paper~\footnote{For a thorough review of TBA, we refer the interested reader to~\cite{takahashi_thermodynamics_1999}.}.

The solutions of the Bethe equations are expressed in terms of quasi-momenta or rapidities, and as one takes the thermodynamic limit, one introduces a density of allowed quasi-momenta $\rho^T(\rap)$, corresponding to a total density of states. In general, we will also introduce an additional discrete quasiparticle label $j$; for XXZ-like spin chains, $j$ is known as the string index. Each allowed quasi-momentum state can be occupied or not in a given macrostate, so one defines the hole density, $\rho^h_j$, and the density of occupied states (quasiparticle density) $\rho_j$, and we have $\rho^T(\rap) = \rho_j(\rap) + \rho^h_j(\rap)$ by definition. The particle density together with hole density fully characterize the state of the system. The expectation value of a conserved quantity in such a state can be written as $Q=\int \rhop_j(\rap) q_j(\rap) d\rap$ where $q(\rap)$ is the single-particle charge eigenvalue of the string of type $j$ with quasi-momentum $\rap$. 

Thermodynamic equilibrium properties can be computed by writing the entropy and energy in terms of $\rhop$ and $\rho^h$, and then maximizing the free energy as usual for a given set of Lagrange multipliers corresponding to a GGE. 
This leads to the so-called Yang-Yang equation~\cite{yang_thermodynamics_1969}. The Yang-Yang equation together with the Bethe equation are enough to fully determine the quasiparticle densities $\rho_j$ and $\rho^h_j$, corresponding to a given GGE state. In the following, we will promote these variables locally by assuming local (generalized) equilibrium. 

\section{Generalized hydrodynamics approach}\label{GHD}

We now formulate a hydrodynamic description of such random integrable systems.
The evolution of chaotic quantum systems from local to global equilibrium is described by the framework of hydrodynamics. In that regime, one imagines chopping off the system into hydrodynamic cells that are big enough to assume equilibrium within each cell, but very small compared to the total system size. This separation of scales allows one to assume local equilibrium, where Lagrange multipliers like temperature or chemical potential are allowed to depend on position and time. 

There is one hydrodynamic equation per conserved quantity in the system -- any other information about the system is ``scrambled'' by the quantum dynamics into non-local entanglement that is not measurable by local observables. For each conserved quantity $Q_n = \sum_x q_n(x)$, we can write a continuity equation
\begin{equation}
\partial_t q_n(x,t) + \partial_x j_n(x,t) = 0,
\end{equation} 
where we restricted ourselves to one spatial dimension. Assuming local equilibrium then leads to a relation $j_n=F_n[ \lbrace q_m \rbrace]$ between the currents $j_n$ and the conserved charges $q_m$ at a given position $x$: this relation is an equilibrium property and gives rise to {\em Euler} hydrodynamic equations that govern ballistic transport properties. More generally, one can perform a gradient expansion of the (expectation value of the) currents in terms of the charges, where contributions to the currents coming from gradient terms $D_{nm} \partial_x q_m$ correspond to diffusive contributions to hydrodynamics (see {\it e.g.}~\onlinecite{spohn2012large}). The diffusion constants $D_{nm}$ are not entirely given by equilibrium properties, and have to be determined by other means such as the Kubo formula, or by using kinetic theory calculations. Once the transport coefficients characterizing the relation between currents and charges are known, hydrodynamics provides a simple set of classical, partial differential equations that govern the non-equilibrium dynamics of the system.

\subsection{Generalized hydrodynamics}
The hydrodynamic framework summarized above is completely general, and it was successfully applied to integrable systems~\cite{castro-alvaredo_emergent_2016,bertini_transport_2016}: the resulting framework is now known as {\em generalized hydrodynamics} (GHD), as it describes systems in local GGE equilibrium.  There, local equilibrium  is characterized by the densities $\rho_{j, \rap}(x,t), \rho^h_{j, \rap}(x,t)$, with the charges
 $q_n(x)=\sum_j \int \rhop_{j,\rap}(x) q_{n,j,\rap}(x) d\rap$ and currents $j_i(x,t)=\int \veff_{i,\rap}(x,t) \rhop_{i,\rap}(x,t) d\rap $ (ignoring gradient (diffusive) corrections). $v_j^{\text{eff}}$ is interpreted as a group velocity which is a functional of the quasiparticle density in general. The continuity equations for the conserved charges then imply a continuity equation for the quasiparticle density~\cite{castro-alvaredo_emergent_2016,bertini_transport_2016}
\begin{equation}
\partial_t \rhop_{j,\rap} + \partial_x(\veff_{j,\rap} \rhop_{j,\rap}) = 0, \label{GHD equation}
\end{equation}
For a non-interacting system, $\veff_j$ is independent of $\rhop_j$. In that case, in clean systems, diffusive corrections to~\eqref{GHD equation} are believed to be absent. 

We now turn to our specific example~\eqref{eq:Free ham in spin}, which is noninteracting in the fermionic language. Recall that in the noninteracting limit, $\eta = \pi/2$ in Eq.\eqref{eq:Random parameters}. 
In this limit, there are two strings $j=1,2$, and their group velocity is simply given by~\cite{PhysRevLett.113.187203,castro-alvaredo_emergent_2016,bertini_transport_2016} 
\begin{equation}
\veff_{j,\rap}=\frac{e'_j(\rap)}{p'_j(\rap)}= q_j\frac{e'_j(\rap)}{2\pi \rhot_{j,\rap}}, \label{eq:velocity}
\end{equation}
where $e_j$ is the quasiparticle energy given by 
\begin{equation}e_j(\rap)=4 J A_j(\rap),\label{tbaenergy} \end{equation}
$p_j$ the quasiparticle momentum,  and the total density of states $\rhot_{j,\rap}=\rhop_{j,\rap}+\rho^h_{j,\rap}$ is given~\cite{takahashi_thermodynamics_1999,essler_integrable_2018} by the Bethe equation (which is particularly simple since the model is non-interacting)
\begin{equation}
\rhot_{j,\rap}=q_j \frac{1}{N}\left(\sum_{k=1}^{N/2}A_j \left(\rap+\frac{4}{\pi}\xi_{2k} \right) + \frac{N}{2}A_j(\rap) \right),\label{rho total}
\end{equation}
for a system of size $N$, with the function $A_j(\rap)$ defined as \begin{equation}
A_j(\rap)=\frac{\pi}{4}\frac{q_j}{\text{cosh}(\pi \rap / 2)} \label{function a}.
\end{equation} 
In these equations, $q_1=1, q_2=-1$. In the interacting case, all these quantities would be ``dressed'' and would become functional of the quasiparticle densities $\rho_j$.

As expected since the model is non-interacting, the group velocity does not depend on the density $\rho$ of the other quasiparticles. However, it does depend on the inhomogeneous variables $\xi_i$. It is thus clear that some kind of averaging over these random variables needs to be done in order to formulate a hydrodynamic theory of this random quantum spin chain. 

\subsection{Averaging and coarse-graining}

In order to average of the random variables $\xi_i$, we go back to the physical picture of hydrodynamics and divide the system into mesoscopic hydrodynamic cells large enough to be in the thermodynamic limit. Let the system length be $L$ and it be divided in $N \gg 1$ sub-cells of size $\Delta x=L/N \gg a$ with $a$ the lattice spacing. For a given disorder realization, the velocity in each hydrodynamic cell is given by~\eqref{eq:velocity}. These velocities depend on our choice of sub-cell division, but as we will see this dependence drops out of the final result. 

Given these velocities we can easily construct the trajectory of a given quasi-particle from its initial position $x=0$. Let us find the time required for a quasiparticle of type $j$ with rapidity $\lambda$, initially at $x=0$ to reach $x=M\Delta x$. It is given by~\footnote{We will momentarily omit the string number and rapidity/momentum subscripts to make the equations more readable.},\begin{equation}
t_x=\Delta x \sum_{i=1}^M \frac{1}{v_i},
\end{equation}
with $v_i$ the velocity the $i^{th}$ cell. We then have using~\eqref{eq:velocity}
\begin{equation}
t_x=\Delta x \frac{2\pi}{e'} \sum_{i=1}^M \rhot_i.
\end{equation}
Since $\rhot$ is a sum of random variables, we can use the central limit theorem to deduce that both $\rhot$ and $t_x$ are Gaussian distributed (provided the hydrodynamic cells are large enough, and $M \gg 1$). This also shows that the result is largely independent of the distribution chosen for the random parameter $\xi$, as long as the central limit theorem is applicable, as is the case for the distributions considered in this paper. 

Thus we have following result: the time taken for a quasiparticle to move over a distance $x$ is Gaussian distributed, with the average time being given by
\begin{equation}
\overline{t_x}= x \frac{2 \pi \overline{\rho^T}}{e'}, 
\end{equation}
where $\overline{\rho^T}$ is the disorder average of $\rho^T$. This quantity is clearly independent from our choice of hydrodynamic cells -- it only depends on $x$, not $M$, or $\Delta x$. The standard deviation reads
\begin{equation}
\sigma[t_x]=\sqrt{\frac{\Delta x}{2 N_{\rm sub}}} \sqrt{x} \frac{2 \pi \sigma\left[A(\lambda+\frac{2}{\eta}\xi) \right]}{e'}, \nonumber
\end{equation}
where $N_{\rm sub}$ is the number of lattice sites inside a cell, and $\sigma[A(\lambda+\frac{2}{\eta}\xi)]$ is the standard deviation of the function defined in equation~\eqref{function a}. Note that $\sqrt{\frac{\Delta x}{2 N_{sub}}}=\sqrt{\frac{\latt}{2}}$ with $\latt$ the lattice spacing, implying that the standard deviation of $t_x$ is also independent of our choice of hydrodynamic cells. Thus we conclude that the distribution of $t_x$ does not depend on the partition of  hydrodynamic cells, and is well defined.

We define the average velocity $\vavg$ via the relation $\vavg \overline{t_x}=x$. This yields
\begin{equation}
\label{eqVelocityAv}
\vavg=\frac{e'}{2\pi \overline{\rho^T}}=\overline{(\veff)^{-1}}^{-1}.
\end{equation}
Note that this is {\em not} the average of $\veff$ over disorder.

\begin{figure}
\center
\includegraphics[scale=0.65]{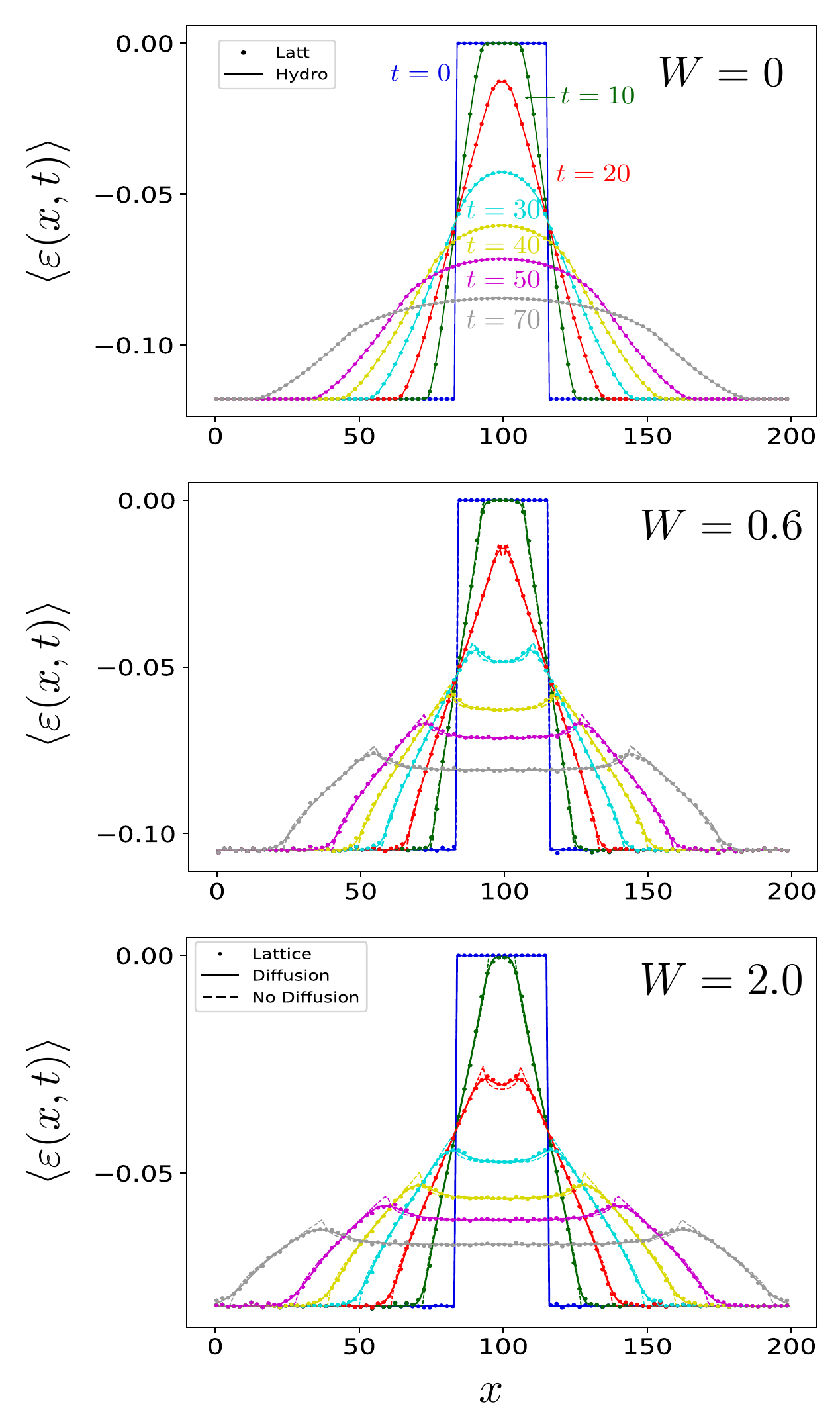}
\caption{{\bf Non-equilibrium energy transport.} The different panels show the evolution of the energy density as a function of time for different disorder strengths, for an initial state at temperature $T=1$ with a small region of the system locally at infinite temperature. We compare exact numerical results, and the hydrodynamic prediction from the solution of  eq.~\eqref{Final PDE}, with and without the diffusive term. As disorder is increased, the diffusive effects become more pronounced, and diffusive corrections are needed to reproduce the numerical data. The numerical data was averaged over $\sim 3 \times 10^3$ and  $\sim 1.6 \times 10^4$ disorder realizations for $W=0.6$ and $W=2.0$, respectively. }
\label{Fig: Thermal energy transport}
\end{figure}

The probability distribution of the time it took for quasiparticle to move over a distance $x$ thus reads
\begin{align}
P_x(t)&=\frac{1}{\sqrt{2\pi \Gamma x}}e^{-(t-x/\vavg)^2/2\Gamma x} \label{Temporal Diffusion},\\
\Gamma &\equiv  \left(\frac{2\pi \sigma[A(\rap +\frac{2}{\eta}\xi)]}{e'}\right)^2{\frac{a}{2}}.
\end{align}
This process is called {\em temporal diffusion}~\cite{boon_temporal_2003,langmann_diffusive_2018}, as it looks like an usual diffusion process where the roles of space and time are exchanged. However, in the hydrodynamic limit, the spreading of the distribution is confined to region $(x/\vavg-t)^2=\mathcal{O}(\Gamma x)$ or $x=\vavg t \left(1+\mathcal{O}(\sqrt{\frac{\Gamma \vavg}{t}}) \right)$. Thus in the limit $\frac{\Gamma \vavg}{t} \ll 1$ we can replace $x$ by $\vavg t$ and get
\begin{equation}
P(x,t) \approx \frac{1}{\sqrt{2\pi \Gamma (\vavg)^ 3 t}}e^{-(x-\vavg t)^2/2\Gamma (\vavg)^ 3 t}, \label{Spatial Diffusion}
\end{equation}
which corresponds to a biased random walk.  A similar temporal diffusion equation recently appeared in the context of energy transport in a random conformal field theory~\cite{langmann_diffusive_2018}. In all the numerical results below, we have checked that the difference between the temporal and ordinary diffusion descriptions are negligible in the hydrodynamic limit. 

For generic initial condition of the quasiparticles, $\rhop^0(x,t=0)$, the evolution should thus reads 
\begin{equation}
\rhop(x,t)=\int \frac{1}{\sqrt{4\pi D t}}e^{-(x-x_0-v^\star t)^2/4D t} \rhop^0(x^0) dx^0, \label{Solution of rhop}
\end{equation}
with $D=\Gamma (v^\star)^3/2$ since the quasiparticles are non-interacting. Reintroducing the string and rapidity labels, we find that the quasiparticle density satisfies the following hydrodynamic equation
\begin{equation}
\boxed{
\partial_t \rhop_{j,\rap} (x,t) + \vavg_{j,\rap}\partial_x \rhop_{j,\rap}(x,t)=D_{j,\rap}\partial_x^2 \rhop_{j,\rap}(x,t),} \label{Final PDE}
\end{equation}
where $D_{j,\rap}\equiv \frac{\Gamma_{j,\rap} (\vavg_{j,\rap})^ 3}{2}$ is a diffusion constant due to the disorder. We emphasize that the transport coefficients $\vavg$ and $D$ in this equation {\em do not} depend on the details of our coarse graining procedure -- in particular, they do not depend on the size of the hydrodynamic cells $\Delta x$ as long as $L \gg \Delta x \gg a$. We emphasize that contrary to diffusive corrections due to thermal fluctuations, the diffusive term in the above equation is diagonal in the quasiparticle basis, and occurs even in the non-interacting case. For interacting random chains, we expect the argument above to carry over for the average velocity~\eqref{eqVelocityAv}, but the diffusion matrix should be more complicated as it will have contributions from interactions and disorder. We leave a detailed analysis of the interacting case for future work.

\section{Non-equilibrium transport} \label{sec:Transport}

We now use the hydrodynamic equation derived above to study non-equilibrium energy and spin transport in the random spin chain~\eqref{eq:Free ham in spin}.  This will also allow us to benchmark and test the validity of the hydrodynamic approach, and investigate the importance of the diffusive terms due to disorder. The energy and spin densities can be expressed in terms of the quasiparticle densities as
\begin{align}
\epsilon(x,t)&=\sum_j \int \rhop_{j,\rap}(x,t) e_{j}(\rap)\ d\rap, \label{eq:Energy}\\
s_z(x,t)&=\frac{1}{2}-\sum_j n_j \int \rhop_{j,\rap}(x,t)\ d\rap, \label{eq:Spin}
\end{align}
where $n_j$ is given in our case by: $n_1=n_2=1$.

\subsection{Energy transport}

We first discuss energy transport. We consider a lattice of $L=200$ sites, prepared it in a thermal state with $\beta = 1$, but with  an interval of $32$ sites in the middle of the system prepared in an infinite temperature state ($\beta=0$). The time evolution of the energy density profile can then be  straightforwardly extracted from the free fermion representation~\eqref{eq:Hamiltonian}. We note that due to next to nearest interactions we have to define energy density on two sites instead of one. We also emphasize that since the initial condition has a mirror symmetry about the middle of the lattice, one expects the evolution to be symmetric as well. However, this is only guaranteed if the sample of random variables $\xi_i$ is symmetric about $\xi=0$ --- this is due to the presence of imaginary interaction in Hamiltonian~\eqref{eq:Hamiltonian} which breaks this mirror symmetry. To retain the mirror symmetry exactly in the numerical data, we have ensured that the sample taken for the random variables are symmetric.

We then compare the numerical results from the solution of the hydrodynamic equation~\eqref{Final PDE}, combined with formula~\eqref{eq:Energy}. Fig.~\ref{Fig: Thermal energy transport} show the hydrodynamic prediction for the energy density (solid line) compared to lattice data, for different disorder strengths $W=0$ (clean case), $W=0.6$ and $W=2.0$. 
We also show the ballistic hydrodynamic prediction ignoring diffusive corrections (dashed line) -- formally setting $D=0$ in the hydrodynamic equation.  The agreement between the numerics and hydrodynamics is excellent, and we find that diffusive corrections are needed to accurately describe the numerical data, especially at stronger disorder.

\subsection{Spin transport}

\begin{figure}
\center
\includegraphics[scale=0.23]{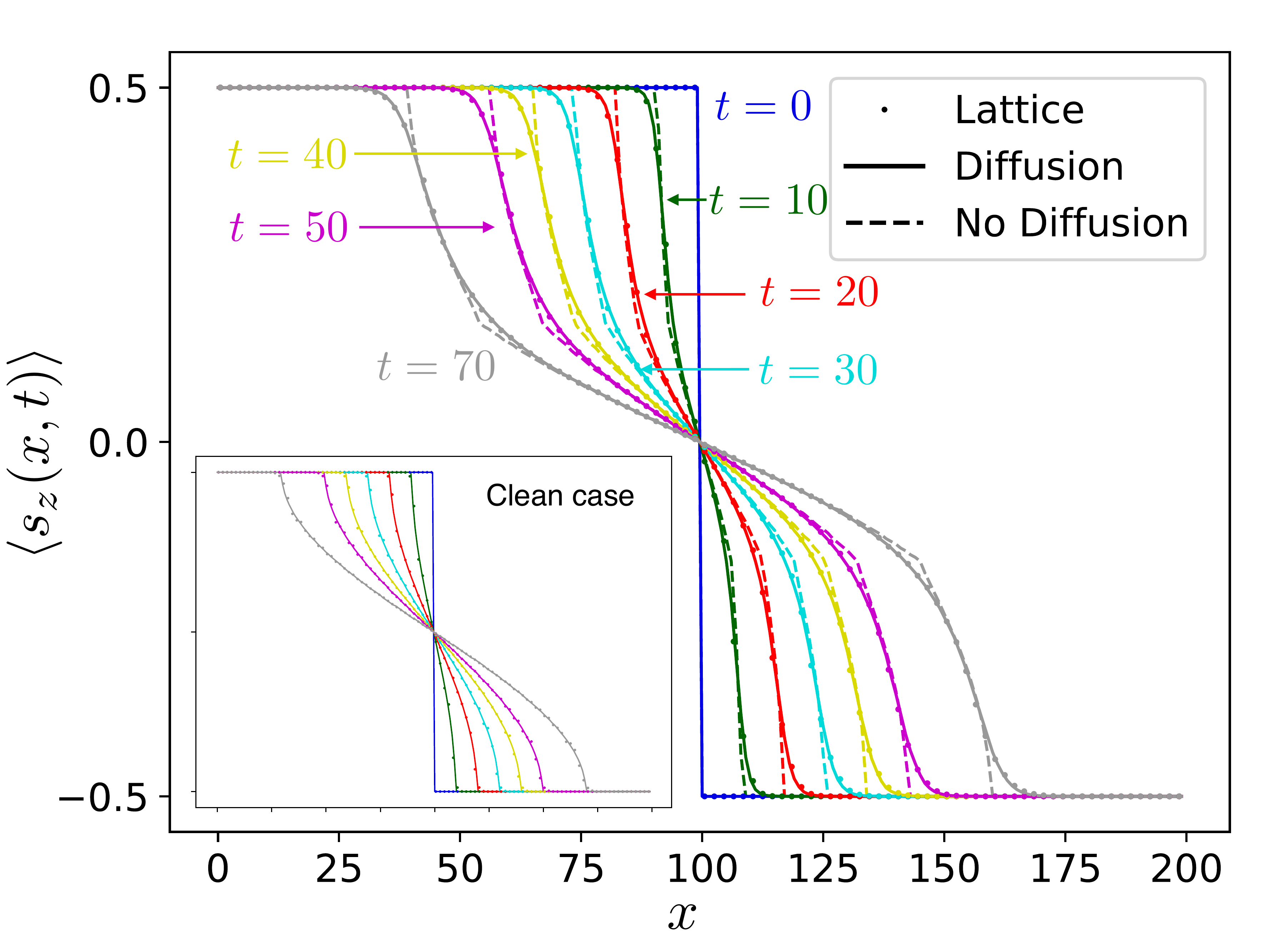}
\caption{{\bf Spin domain wall initial state}. Comparison of the hydrodynamic prediction~\eqref{Final PDE} with numerical results a for spin domain wall initial state. The disorder strength is $W=0.6$, and numerical results are averaged over $\sim 2\times  10^3$ disorder realizations. The hydrodynamic equation including diffusive terms (solid line) is describing the numerical results much more accurately than the purely ballistic prediction (dashed line). This establishes the presence of diffusive terms in the hydrodynamic description of this non-interacting system, even for an initial state that does not incorporate thermal fluctuations. }
\label{Fig:DW}
\end{figure}

We also consider spin transport starting from an initial domain wall state $\Ket{\psi_0} = \Ket{\uparrow \uparrow \dots \uparrow \downarrow \dots \downarrow \downarrow}$ (see Ref.~\onlinecite{PhysRevB.97.081111} and references therein).  This initial state has been considered for clean XXZ spin chains in several recent works:
it leads to a steady-state with zero entropy, where diffusive terms due to interactions and thermal fluctuations are expected to vanish. The hydrodynamic prediction for the local magnetization and numerical data are compared in Fig.~\ref{Fig:DW}. Here also, eq.~\eqref{Final PDE} provides an excellent description of the numerics and our data clearly shows the presence of diffusion in this non-interacting setting where no thermal fluctuations are present~\cite{bulchandani_subdiffusive_2018}. We expect this initial state to be also useful to extend our approach to interacting random spin chains, as it would allow one to isolate diffusive terms due to from disorder (since the diffusion matrix coming from interactions vanishes). Comparing the predictions from hydrodynamics to numerical results in the interacting case would however be very numerically demanding because of the average over disorder, and is left for future work.

\section{Properties of low-energy quasiparticles}
\label{SecWavefunctions}

So far, our discussion has focused on macroscopic energy and spin transport, which is dominated by fast quasiparticles. In addition to these typical, fast, quasiparticles, however, these models also have slow quasiparticles at energy $|E| \approx 0$. These are important, e.g., for low-temperature transport, as well as for the behavior of local autocorrelation functions at late times. We discuss the nature of these quasiparticles here. We first explore the properties of wavefunctions, both numerically and analytically, and find that these undergo a quasi-localization transition for the exponential disorder distribution~\eqref{expdis}. We then apply the thermodynamic Bethe ansatz results in Sec.~\ref{GHD} to study the asymptotic behavior of the velocity and density of states as $|E| \rightarrow 0$. Our discussion here is confined to the non-interacting model~\eqref{eq:Free ham in spin}. 

\subsection{Spatial correlations of eigenstates}

\begin{figure}[tb]
\begin{center}
\includegraphics[width = 0.42\textwidth]{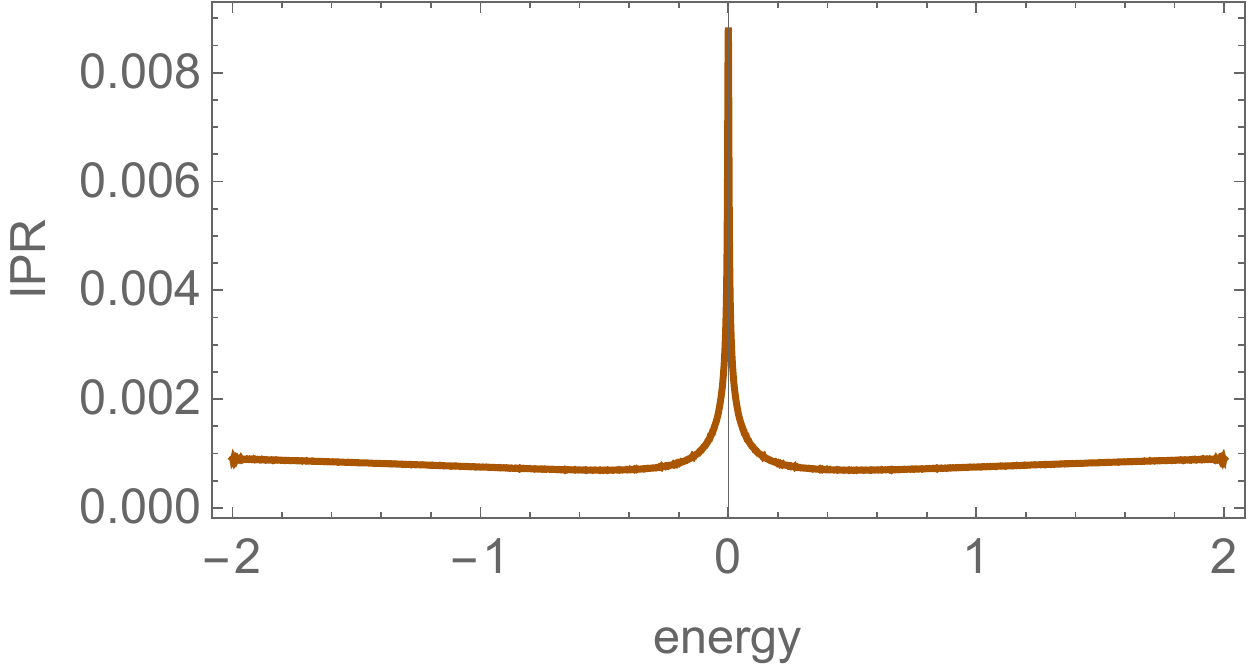}
\includegraphics[width = 0.42\textwidth]{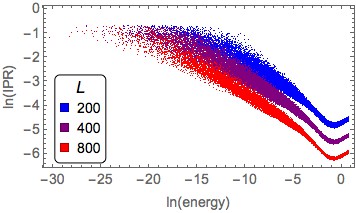}
\caption{{\bf Low-energy wavefunctions.} Upper panel: IPR vs. energy for the non-interacting model~\eqref{eq:Free ham in spin} with Gaussian-distributed disorder with parameters $W = 1$, $L = 2000$ (single, typical, realization). Lower panel: Log-log plot of IPR vs. energy for an exponential disorder distribution, with $\phi = 0.5$, binned over 500 realizations for each system size.}
\label{iprfig}
\end{center}
\end{figure}

We first discuss the properties of single-particle eigenstates. Because of the Bethe-ansatz integrability of the model, eigenstates are never localized in the conventional sense, and most are completely delocalized~\cite{essler_integrable_2018}. However, the states closest in energy to zero have anomalous properties, which show up in the hydrodynamic framework as very slow velocities. We now address these properties more directly. Fig.~\ref{iprfig} shows that the inverse participation ratio (IPR) $\sum_x \left| \psi_x \right|^4$ rises steeply for states very near zero energy; thus, these are less spread out than the rest of the spectrum, which is fully delocalized. In the rest of this section we will focus on the few relatively localized states near zero energy. 

\begin{figure}[!t!b]
\begin{center}
\includegraphics[width = 0.42\textwidth]{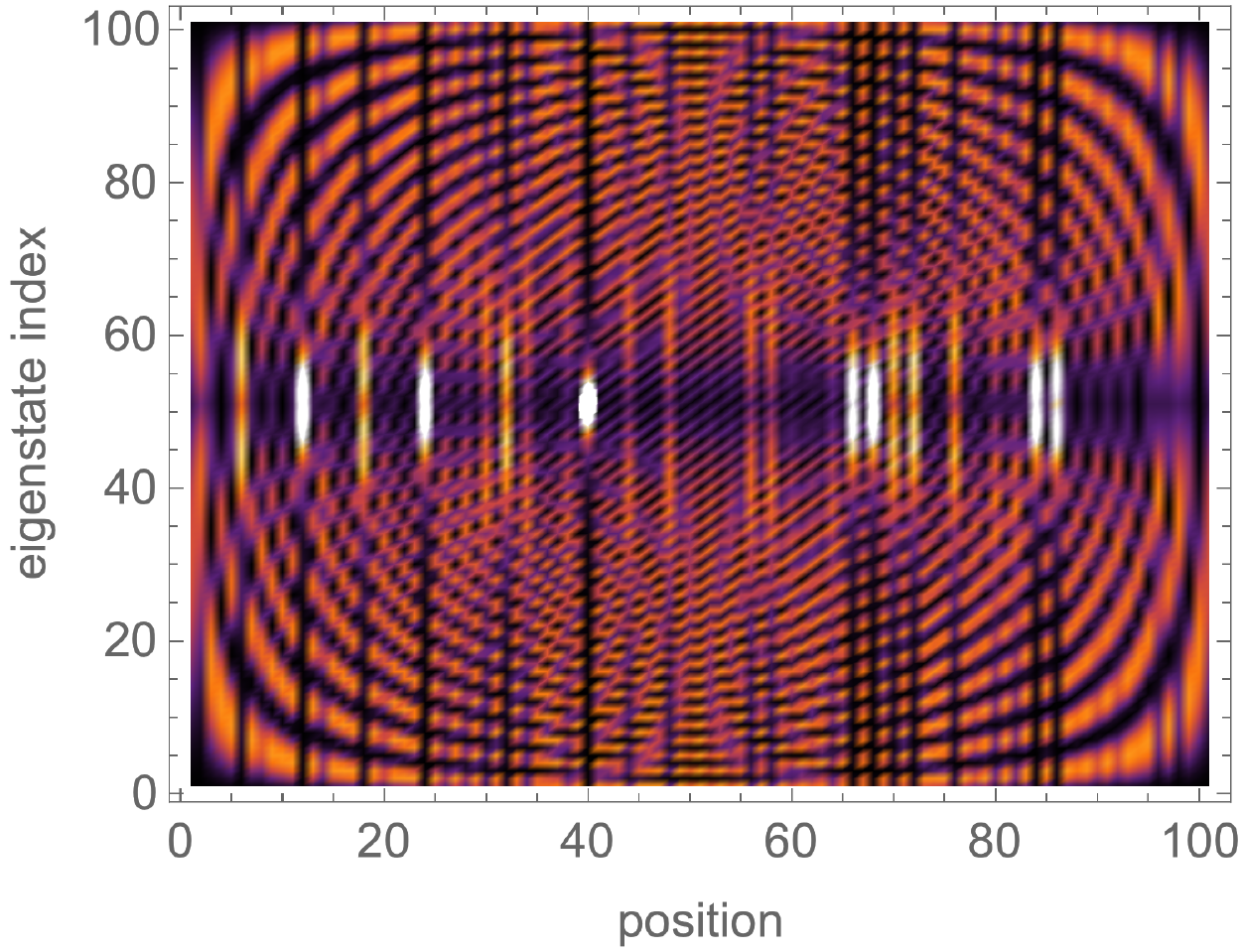}
\includegraphics[width = 0.42\textwidth]{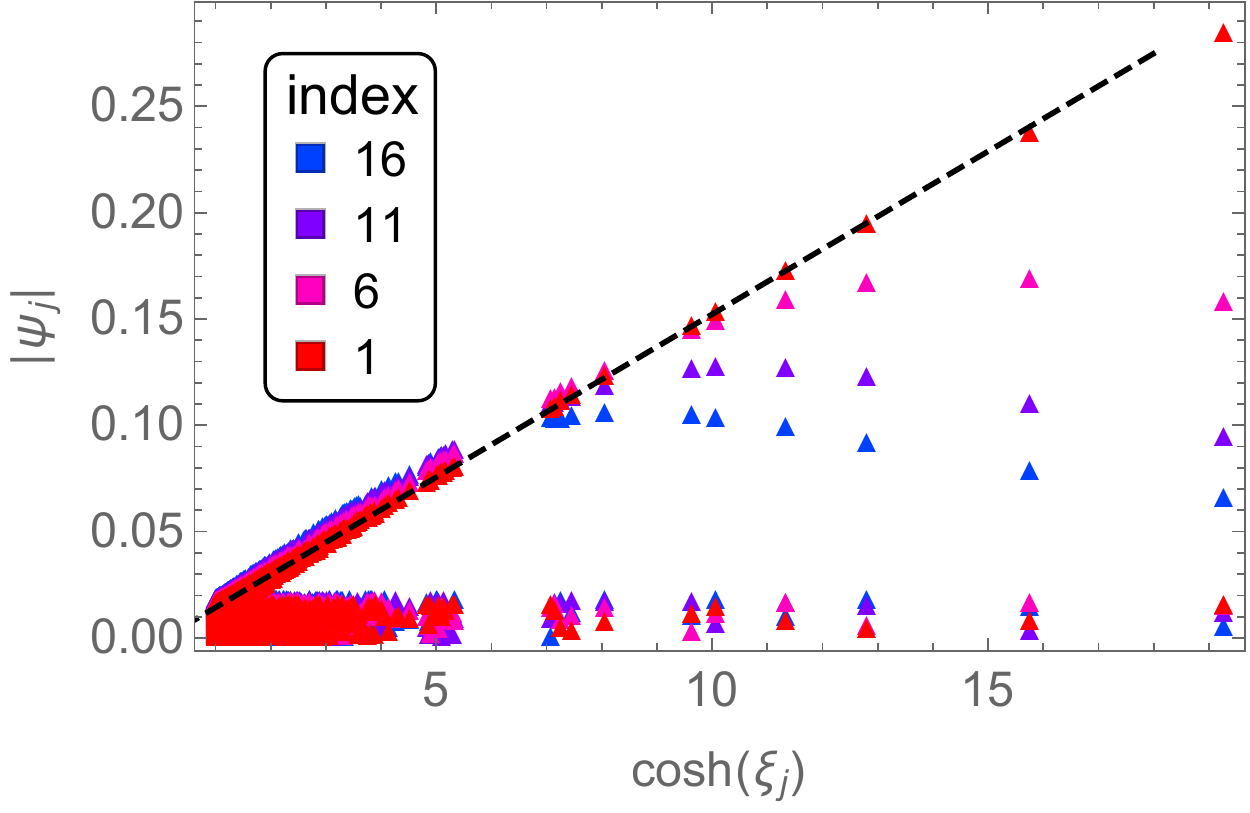}
\caption{{\bf Spatial structure of low-energy wavefunctions.} Upper panel: density plot of eigenstate probability density in space for a single eigenstate in a single realization at $L = 100$; $y$ axis is the eigenstate index, a proxy for the energy. All the states near zero energy are concentrated at the same sites, i.e., they are perfectly spatially correlated. Lower panel: scatterplot of eigenstate amplitude vs. inverse hopping matrix element $\cosh(\xi_j)$ for the anomalous eigenstates for $L = 2000$. The figure shows four eigenstates; states with the smallest $|E|$ are shaded red while those with larger $|E|$ are shaded blue.}
\label{spatialcorr}
\end{center}
\end{figure}

Given the anomalously large IPRs of states near zero energy, it is natural to investigate their spatial structure. It turns out this structure is simple and striking. The upper panel of Fig.~\ref{spatialcorr} plots the spatial profile of each eigenstate against the eigenstate index. States near $|E| = 0$ are halfway up the $y$ axis; they are evidently strongly peaked at certain lattice sites, and the peak locations are the \emph{same} for each such state. In other words, these states exhibit a very strong form of Chalker scaling~\cite{PhysRevLett.61.593,CHALKER1990253}. Further, the peak locations and intensities follow a simple pattern (lower panel of Fig.~\ref{spatialcorr}): the amplitude of each peak is $\propto \cosh(\xi_j)$, where $\xi_j$ is the nearest neighbor hopping at the peak. These observations can be qualitatively understood in terms of a simple picture in which the state has a roughly uniform current, and to maintain this uniform current the density piles up near weak links. Thus some form of ``hydrodynamics'' applies even at the level of individual wavefunctions in individual realizations.
As one moves away from zero energy, the states lose intensity first at the strongest maxima (Fig.~\ref{spatialcorr}); higher-energy wavefunctions bypass these weak links instead of piling up near them.

We find, numerically, that these anomalous states are supported entirely on even sites (i.e., sites for which the next-nearest neighbor coupling is absent). This is intuitively plausible since no value of $\xi_j$ can simultaneously suppress both the nearest-neighbor and next-nearest-neighbor hopping out of an odd site.

\subsection{Multifractality and quasi-localization}

\begin{figure}[tb]
\begin{center}
\includegraphics[width = 0.43\textwidth]{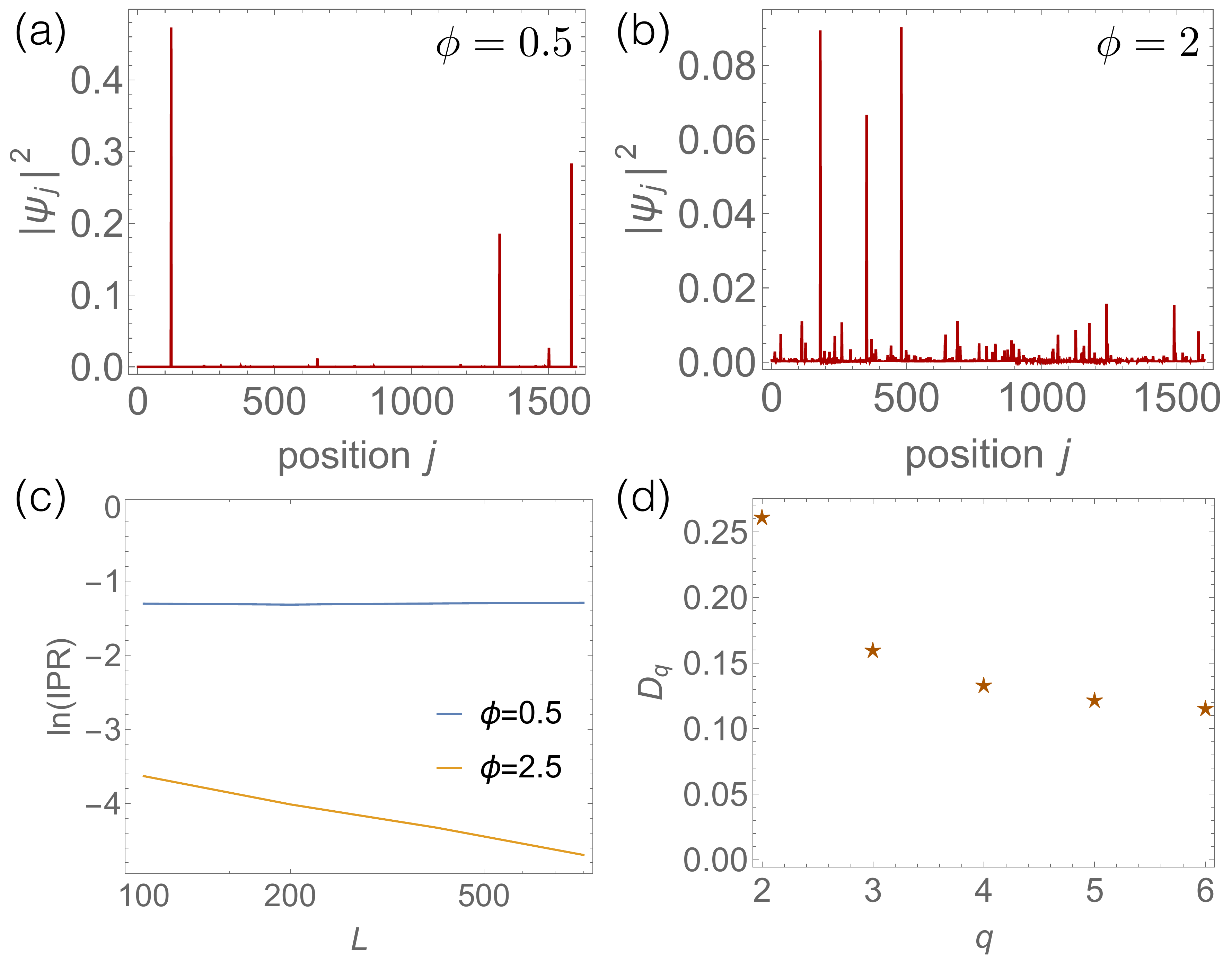}
\caption{ {\bf Properties of the noninteracting model with an exponential distribution of $\xi_j$}  (a), (b)~Spatial structure of single low-energy eigenstates for $\phi = 0.5$ (quasi-localized) and $\phi = 2$ (multifractal), for system size $L = 2000$. (c)~Decay of the IPR with system size in the two phases, averaged over 1000 samples at each size. (d)~Multifractal exponent $D_q$ (where $q$ labels moments of the wavefunction) for $\phi = 2$. The exponents are extracted for system sizes from $L = 100$ to $L = 800$ with 1000 samples per size. For conventional delocalized states, such as the states away from zero energy, one would have $D_q = 1$ for all $q$.}
\label{multifractality}
\end{center}
\end{figure}

The eigenfunctions near zero energy are sharply peaked at a few (even) sites, suggesting that they might be critical rather than conventionally delocalized. 
The nature of these states is highly sensitive to the tails of the disorder distribution. 
So far we have considered Gaussian distributions of the $\xi_j$; numerical extraction of the IPR at the accessible system sizes does not settle whether the wavefunctions are localized or critical. 
However, the relationship between the wavefunction amplitudes and $\cosh(\xi_j)$ allows us to address this question semi-analytically. Among $L$ Gaussian random variables, the largest value is likely to be $\exp(-\xi^2 / W^2) \simeq 1/L$, so $\xi = W \sqrt{\log L}$. The (unnormalized) wavefunction amplitude at this site is therefore $\cosh(W \sqrt{\log L}) \sim \exp(W \sqrt{\log L})$. Meanwhile, the typical value of $\cosh(\xi_j)$ is of order unity, and occurs $O(L)$ times. Thus the weakest link in a typical sample does not affect its properties overall. We therefore expect that the low-energy wavefunctions in the Gaussian case are asymptotically not multifractal. 

This reasoning also suggests that to get multifractal wavefunctions, it suffices to change the disorder distribution from Gaussian to exponential, $P(\xi) \propto \exp(-\phi \left| \xi \right|)$. In this case, repeating the argument in the previous paragraph gives that the largest typical peak has amplitude $L^{1/\phi}$. When $\phi < 2$ the wavefunction piles up at the weakest link, and is ``quasi-localized'' in the sense that its IPR is independent of system size. Numerical simulations clearly show this quasi-localized behavior at small $\phi$, as well as the expected multifractal behavior at larger $\phi$ (Fig.~\ref{multifractality}(d)). We characterize multifractality through the quantity~\cite{RevModPhys.80.1355}
\beq
D_q = \frac{1}{(1-q) \log L} \overline{\log \sum\nolimits_x  |\psi_x|^{2q} },
\eeq
where $\overline{O}$ refers to the disorder average of the quantity $O$. For each sample, we take the lowest-$|E|$ state. 

\begin{figure}[tb]
\begin{center}
\includegraphics[width = 0.4\textwidth]{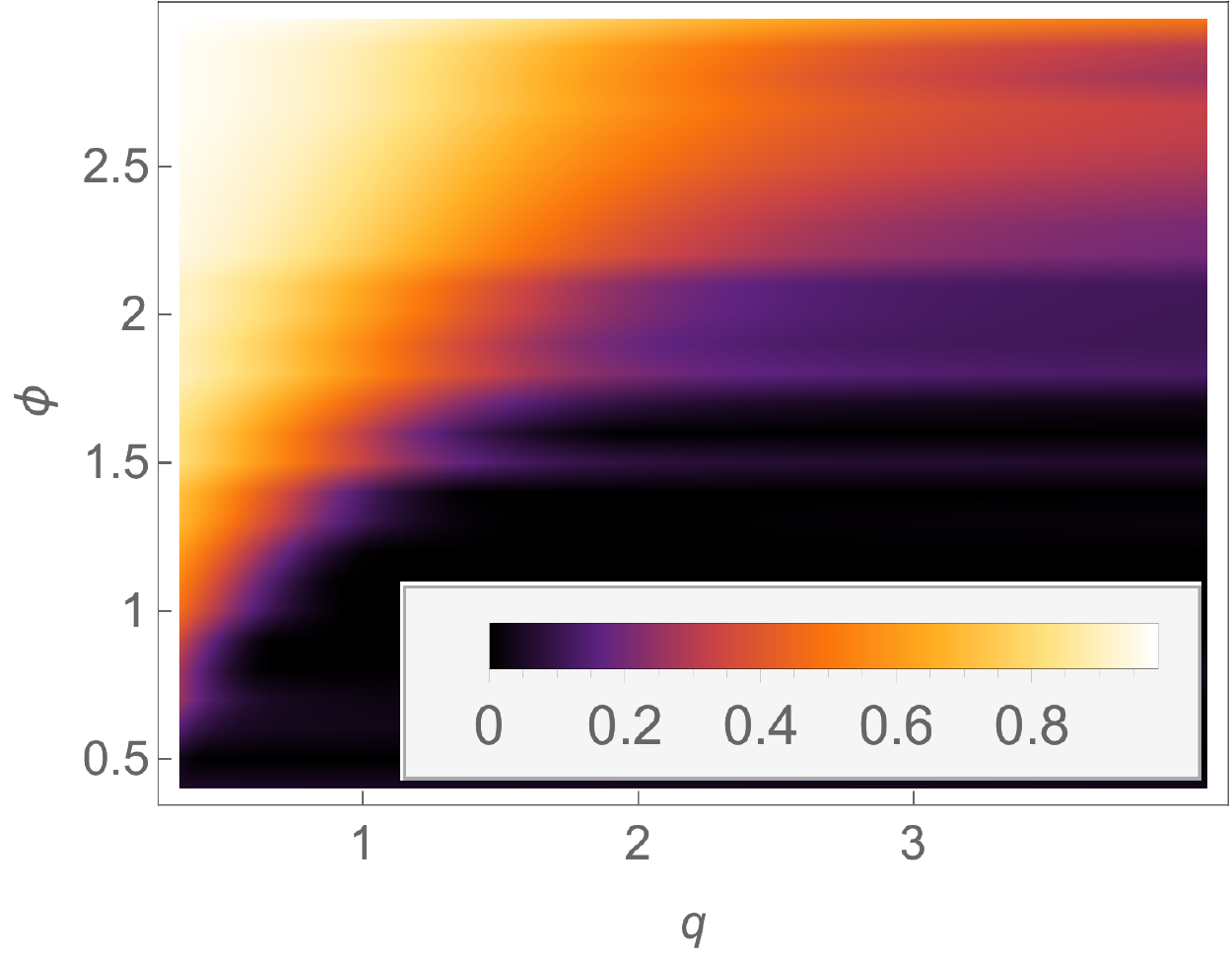}
\caption{{\bf Multifractal spectrum.} Density plot of the quantity $D_q$ as a function of $q$ and the inverse disorder strength $\phi$ for an exponential distribution. The region in black for $\phi < 2$ is quasi-localized.}
\label{iprvsq}
\end{center}
\end{figure}

In the quasi-localized regime, $D_q = 0$ for $q > 1$. Strictly speaking, this statement is asymptotically true, in the following sense: wavefunctions at any fixed energy away from zero are delocalized, but as one approaches zero energy their IPR approaches a fixed, size-independent value (Fig.~\ref{iprfig}). Thus, there is no mobility edge. 
Although the IPR is finite, unlike an Anderson insulator, the quasi-localized phase has $D_q > 0$ for $q < 1$: although the wavefunction has a finite fraction of its weight concentrated in a few peaks, the rest of the weight is spread out evenly rather than falling off exponentially away from the peaks (Fig.~\ref{iprvsq}). To our knowledge this quasi-localization property has not been previously noticed. 
Finally, we remark that although its IPR is independent of system size, this does not imply that a particular state will be unaffected by adding sites to one end of the system: the wavefunction will remain very sharply peaked at the weakest link, but the location of this link will move as sites are added to the system. 

\subsection{GHD approach to quasi-localization}\label{ghdql}

We now use the results in Sec.~\ref{GHD} to consider the behavior of quasiparticles in the $|E| \rightarrow 0$ limit. From Eq.~\eqref{tbaenergy} we see that these quasiparticles correspond to large $|\lambda|$: in fact, $e_j(\lambda) \sim e^{-\pi|\lambda|/2}$. In what follows we suppress the index $j$ and denote the energy as $E$. The limiting behavior of the velocity and the density of states is sensitive to the tails of the disorder distribution. For the sample-averaged density of states, Eq.~\eqref{rho total} yields
\beq\label{expval0}
\rhot_{1, \rap} = \left\langle \frac{\pi}{8\cosh(\pi \lambda/2 + 2\xi)} \right\rangle_{\rm dis} + \frac{\pi}{8 \cosh(\pi\lambda/2)},
\eeq
where $ \left\langle \dots \right\rangle_{\rm dis} $ denotes the average over disorder.
If the disorder is bounded or falls off faster than exponentially, one can safely approximate $\cosh(x) \approx e^x / 2$ for large enough $\lambda$. The quasiparticle velocity (given by Eq.~\eqref{eq:velocity}) therefore remains nonzero in the $\lambda \rightarrow \infty $ limit, so transport is asymptotically ballistic (but with a slower velocity). Likewise, the density of states remains finite. To find the density of states, one notes that the number of states in a rapidity interval $\delta \lambda$ is given by $\rhot_{1, \rap} \delta \lambda$. These cover an energy window $\delta E = E'(\lambda) \delta \lambda = e^{-\pi \lambda/2} \delta \lambda$. Thus, $\rho(E) = \rhot_{1,\rap} / E'(\lambda)$, which remains finite for Gaussian distributed disorder in the $|\lambda| \rightarrow \infty $ limit (although in practice the suppression is quantitatively quite large).

\begin{figure}[tb]
\begin{center}
\includegraphics[width = 0.45\textwidth]{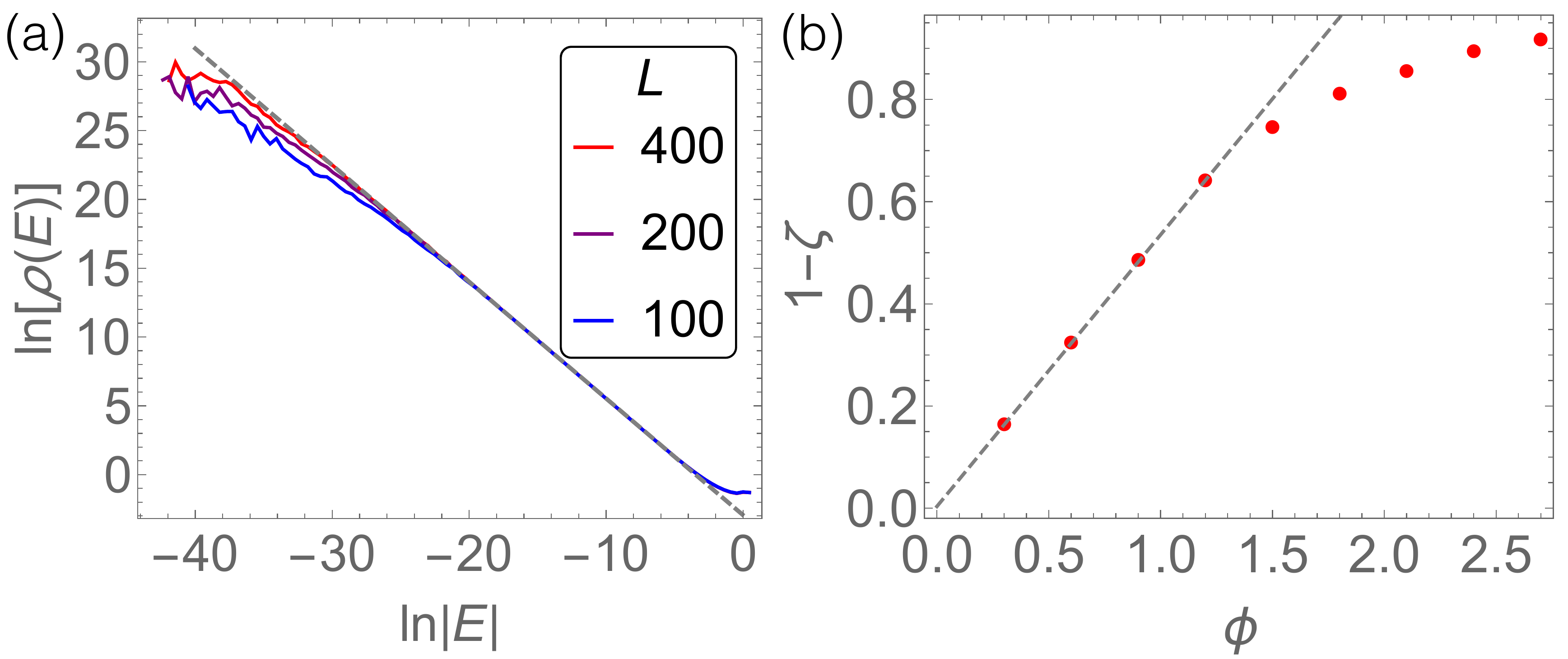}
\caption{\textbf{Density of states and quasi-localization}. (a)~Log-log plot of density of states for various system sizes, in the case where $\xi_k$ are distributed exponentially with the parameter $\phi = 0.3$. The density of states diverges as $|E|^{-\zeta}$ at low energies. (b)~Exponent $1 - \zeta$ vs. $\phi$. For small $\phi$ we find a good linear fit to $\zeta = 1 - 0.53 \phi$, which is in reasonable agreement with the Bethe ansatz prediction $\zeta = 1 - \phi/2$ (see main text).}
\label{figdos}
\end{center}
\end{figure}

However, for distributions $P(\xi)$ with exponential or slower tails, computing the expectation value~\eqref{expval0} is more subtle. We focus on the exponential case $P(\xi) = \frac{1}{2\phi} \exp(-\phi |\xi|)$, which was discussed numerically above. In this case, the expectation value~\eqref{expval0} reads
\beq
\frac{\phi \pi}{4} \int_{-\infty}^\infty d\xi \frac{e^{-\phi |\xi|}}{\cosh(\pi\lambda/2 + 2\xi )}.
\eeq
To get compact expressions for the asymptotics, we approximate $\cosh(\pi\lambda/2 + 2\xi) \approx \frac{1}{2} e^{|\pi\lambda/2 + 2 \xi |}$. There are two cases. When $\phi > 2$, the integral is dominated by small $|\xi|$ and we get
\beq
\left\langle \frac{\pi}{8\cosh(\pi \lambda/2 + 2\xi)} \right\rangle_{\rm dis} \approx \frac{\phi \pi e^{-\pi\lambda/2}}{2 (\phi - 2)} (1 - e^{-(\phi - 2) \pi\lambda/2}) + \ldots
\eeq
where $\ldots$ indicates terms that do not become singular in the limit $\phi \rightarrow 2$. Thus the zero-energy density of states and velocity remain finite when $\phi > 2$, but respectively diverge and vanish as $\phi \rightarrow 2^+$. Note that there are nonanalytic corrections to the density of states, even in this regime: specifically, $|\rho(E) - \rho(0)| \sim E^{\phi - 2}$. 

In the opposite limit $\phi<2$, Eq.~\eqref{expval0} is dominated by $|\xi| \approx \pi\lambda/4$. In this case we have instead 
\beq
\left\langle \frac{1}{\cosh(\pi\lambda/2 + 2\xi)} \right\rangle_{\rm dis} \sim e^{-\phi \pi \lambda / 4}.
\eeq
The velocity then vanishes as $v(\lambda) \sim \exp[-(\pi\lambda/2) (1 - \phi /2)]$, or equivalently $v(E) \sim |E|^{1 - \phi /2}$. Correspondingly the density of states $\rho(E)$ diverges as 
\beq
\rho(E) \sim |E|^{-1 + \phi /2}. 
\eeq
This behavior of the density of states is borne out numerically (Fig.~\ref{figdos}). 

Thus, the dispersion relation of elementary excitations within generalized hydrodynamics agrees with our simple counting estimate in Sec.~\ref{SecWavefunctions}: a transition occurs when $\phi = 2$. When $\phi > 2$ the velocity approaches a finite value as $|E| \rightarrow 0$, so quantities such as the local autocorrelation function behave in an asymptotically ballistic fashion. On the other hand when $\phi < 2$ the velocity vanishes as $|E| \rightarrow 0$, and the local autocorrelation function will in general be anomalous. It is interesting to note that despite the very local character of the rare low-energy states due to  almost disconnected sites, they appear naturally as slow quasiparticles in the hydrodynamics framework.

\section{Operator Spreading and Local Autocorrelations}
\label{sec:OTOC}

\begin{figure}
\center
\includegraphics[width = 0.45\textwidth]{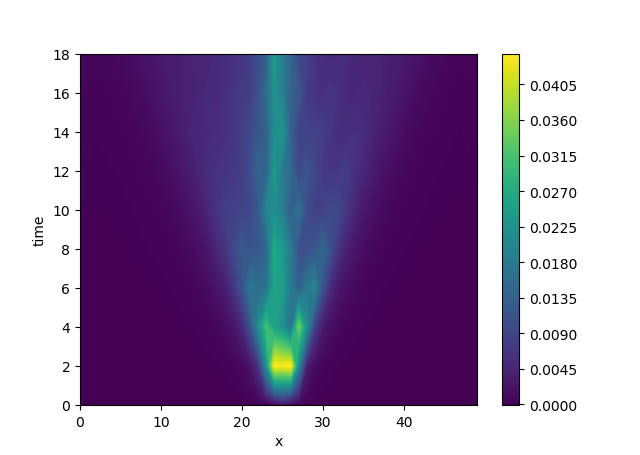}
\caption{{\bf Operator spreading. } Contour plot of the OTOC $C(x,t) = \frac{1}{2}\left\langle [n_x(t), n_0(0)]^2\right\rangle$ for disorder strength $W=1$ and temperature $T=1$,  averaged over 100 realizations. The ballistic spreading of operators and the broadening of the front with time are clearly visible, as well as quasi-localized quasiparticles due to the anomalous low-energy properties of this model.}
\label{Fig: OTOC Contour}
\end{figure}

\subsection{Front broadening}

We close this paper by discussing the consequences of our results for operator spreading in random integrable systems. The dynamics of operator spreading has attracted a lot of attention in recent years, due to its possible connection to many-body quantum chaos~\cite{lo_otoc, ShenkerStanfordButterfly, maldacena2016bound, tsunami, casini2016, mezei2017, FawziScrambling, AdamCircuit1, nvh, vrps, kvh, rpv, cdc_long,PhysRevLett.121.264101,lin_out--time-ordered_2018,2018arXiv180200801X}. Under unitary dynamics, initially local operators spread in space ballistically (unless the system is many-body localized), with an operator ``front'' or light-cone that generically broadens diffusively as ~$t^{1/2}$ in one dimensional chaotic (non-integrable) quantum systems. This is in sharp contrast with (clean) non-interacting systems have an operator front that broadens as $t^{1/3}$ ~\cite{fagotti_higher-order_2017}, governed by an Airy kernel. However, {\em interacting} integrable systems have been argued to have a front that  also broadens as $t^{1/2}$ ~\cite{sg_ffa, gopalakrishnan_hydrodynamics_2018}, just like chaotic systems. (With the notable exception of non-generic zero entropy initial states like the spin domain wall initial state discussed above~\cite{bulchandani_subdiffusive_2018}.) In integrable systems, the operator front is governed by the fastest quasiparticle in the system~\cite{gopalakrishnan_hydrodynamics_2018}: from our hydrodynamic description of random integrable spin chains, it is natural to expect this front to broaden diffusively even in the non-interacting case, since the fastest quasiparticle follows a biased random walk. For these systems, the diffusive broadening of the operator front is due to the local disorder which causes the fastest quasiparticle (on average) to ``wiggle'' around its average trajectory.  

\begin{figure}[tb]
\begin{center}
\includegraphics[width = 0.45\textwidth]{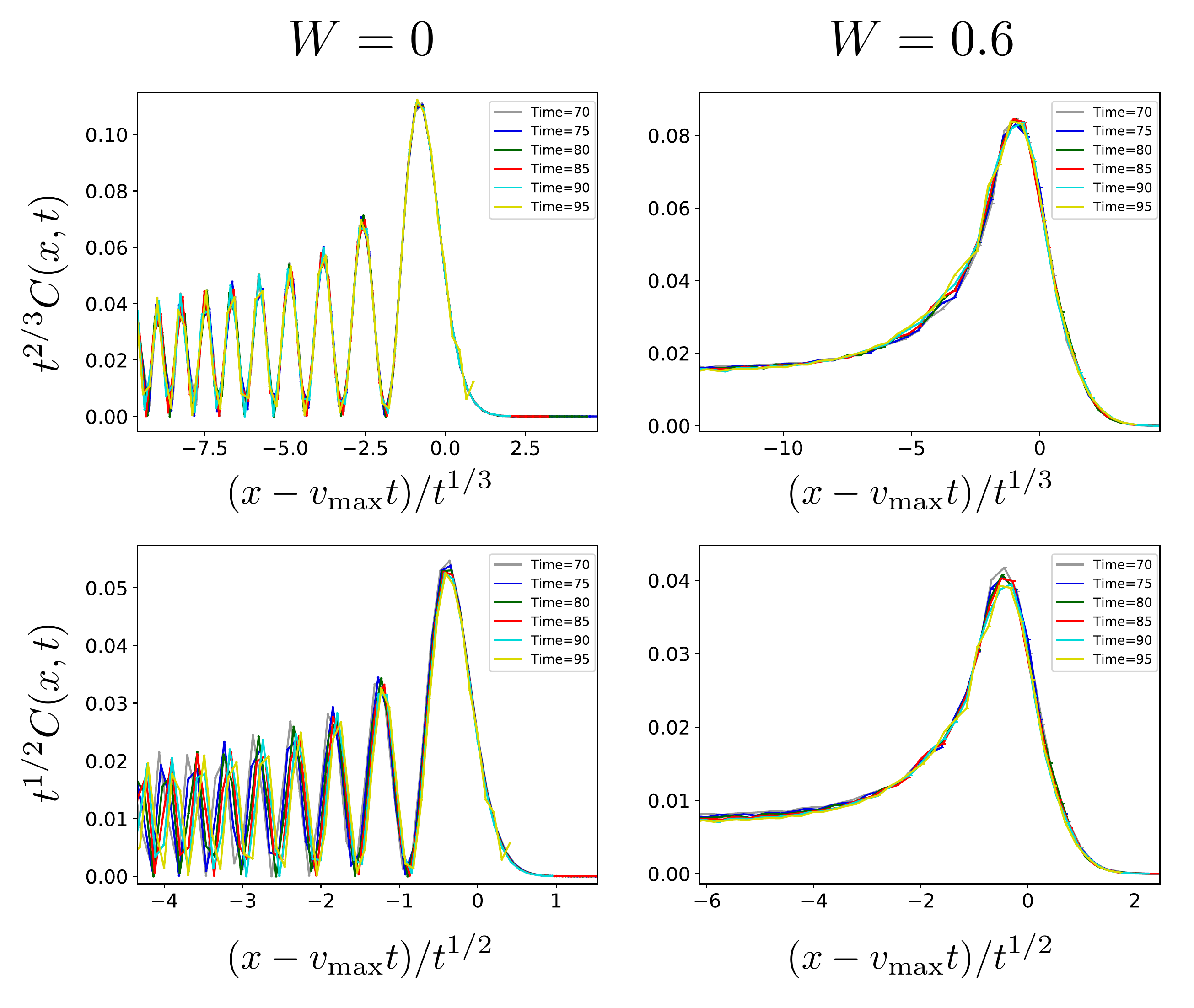}
\caption{{\bf Scaling of the operator front. } Scaling collapses of the OTOC near the operator front. \emph{Left panel:} For the clean case, the front has oscillations as expected from an Airy kernel. The scaling with exponent $1/3$ (upper left) leads to a better collapse than the diffusive scaling $1/2$ (lower left). \emph{Right panel:} Similar collapses for the random case are inconclusive.}
\label{Fig:OTOC}
\end{center}
\end{figure}

In order to characterize operator spreading in our system, we compute a specific out of time ordered commutator (OTOC)~\cite{lo_otoc, ShenkerStanfordButterfly, maldacena2016bound}
given by $C(x,t) = \frac{1}{2}\left\langle [n_x(t), n_0(0)]^2\right\rangle$ in a given thermal state, with $n_x = c_x^\dagger c_x$ the particle density in the fermionic language. Fig.~\ref{Fig: OTOC Contour} shows $C(x,t)$ averaged over 100 disorder realizations for $W=1$ and temperature $T=1$, where a clear ballistic light cone can be observed. We also remark that the OTOC shows significant weight that remains near $x=0$ even after a long time. This behavior has to do with the anomalous low-energy states (Sec.~\ref{SecWavefunctions}), and we will return to it below.   

We now focus on the operator front (light-cone), and investigate whether it broadens as $t^{1/3}$ as expected for clean non-interacting systems, or as $t^{1/2}$ due to diffusive effects. 
We start with the clean case (XX model), and show scaling collapses of the OTOC near the front. It is clear from Fig.~\ref{Fig:OTOC} that the data collapses almost perfectly using a $t^{1/3}$ ansatz. This is consistent with expectations in the clean case, the OTOC shows oscillations characteristic of an Airy Kernel associated with these higher-order corrections to hydrodynamics~\cite{fagotti_higher-order_2017}. In the random case, we expect that these oscillations should vanish, and that the front will collapse onto a diffusive form. From Fig.~\ref{Fig:OTOC}, it is clear that there are no characteristic oscillations near the front. However, scaling our data with both $t^{1/2}$ and  $t^{1/3}$ leads to equally good collapses, and suggest that such collapses are not a very conclusive way to measure the exponent $\alpha$ of the front broadening $t^\alpha$. We also considered collapses from the time evolution following local quenches, with very similar results. Nevertheless, our transport results combined with our hydrodynamic theory strongly suggest that $\alpha=1/2$ even for random non-interacting chains. 

\subsection{Slow local relaxation}

We now turn briefly to the part of the local operator that remains near its initial position at late times. Since we are considering a noninteracting model, we can equivalently consider the return probability of an initially local wavepacket~\cite{sg_huising}. We should distinguish between average and typical behavior: \emph{typically} the initial site is not a weak link, so anomalous wavefunctions have no support there. However, the site-averaged local autocorrelation function does receive a contribution from rare sites. Once again, we discuss this for the exponential distribution~\eqref{expdis} in the quasi-localized phase, $\phi < 2$. 

\begin{figure}[tb]
\begin{center}
\includegraphics[width = 0.44\textwidth]{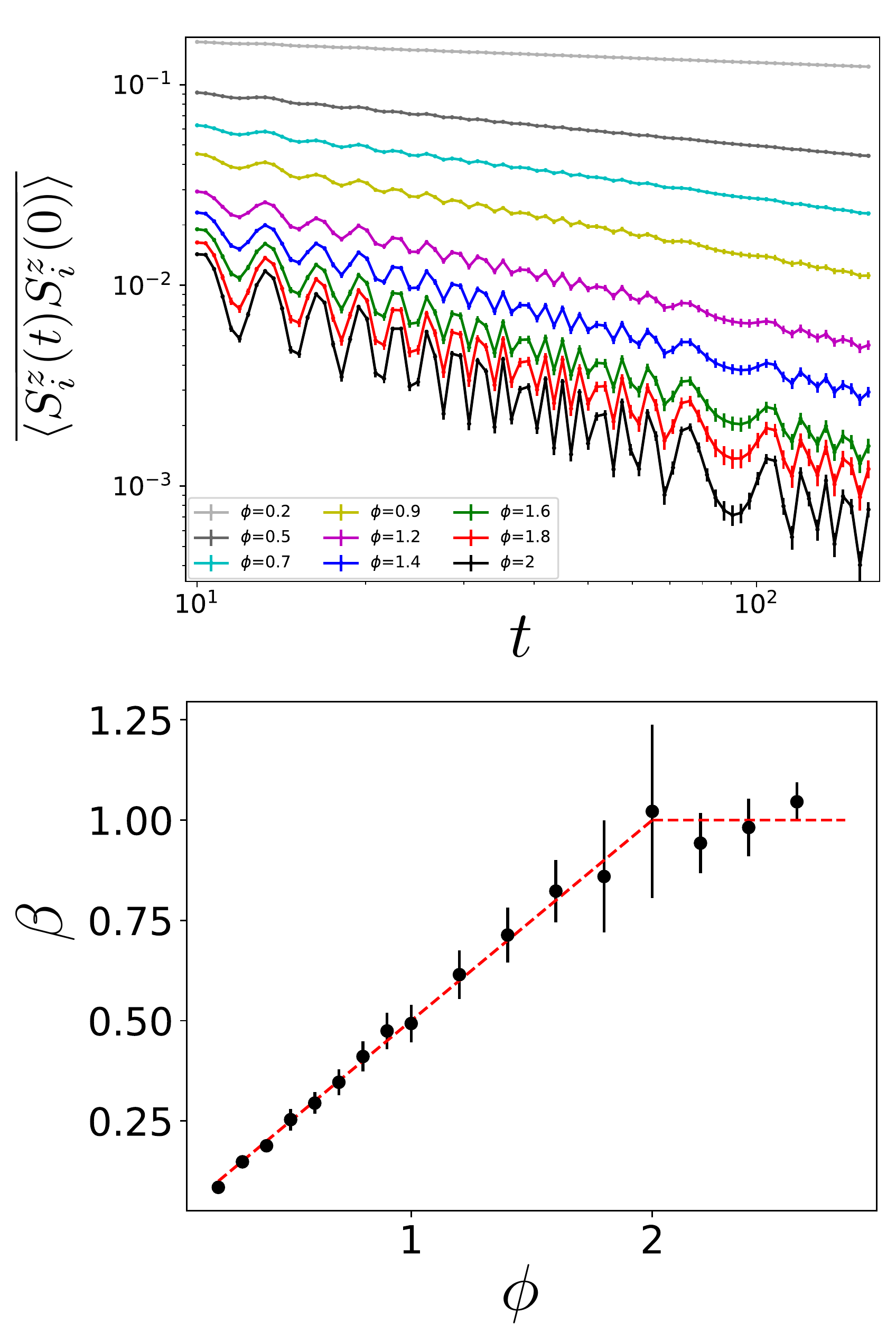}
\caption{{\bf Anomalous local relaxation.} {\it Top panel}:  Algebraic decay of the average local structure factor $C_0(t)$ as a function of time, for various disorder strengths $\phi$. This power-law decay for small values of $\phi$ can be observed up to very long times.
{\it Bottom panel}: 
Decay exponent of the average local correlation function $C_0(t)$, as a function of disorder strength $\phi$. We find that $C_0(t) \sim t^{-\beta}$, where $\beta \approx \phi/2$ throughout the quasi-localized phase $\phi < 2$. The generalized hydrodynamics prediction $\beta = \phi/2$ is indicated by a dashed red line.  When $\phi > 2$ one has conventional ballistic behavior $C_0(t) \sim 1/t$.}
\label{return}
\end{center}
\end{figure}

We consider the probability that a particle initially localized at site $x$ has moved a distance less than one lattice site at time $t$. This quantity is proportional to the (mean) local autocorrelation function $C_0(t)=\overline{\langle S^z_i(t) S^z_i(0) \rangle}$. We focus on $i$ even and infinite temperature. 
In generalized hydrodynamics this can be expressed as
\beq\label{c0t}
C_0(t) \sim \int d\lambda \langle \rho(\lambda) \Theta(a - |v(\lambda) t|) \rangle_{\mathrm{dis.}}, 
\eeq
where at infinite temperature for a free-fermion model, $\rho^T \sim \rho$, and $a=1$ is the lattice spacing. Focusing on low-energy quasiparticles, this integral can be written out as
\beq
C_0(t) \sim \int d\lambda \, d\xi \, e^{-\phi|\xi|} \rho_t(\lambda, \xi) \Theta(1 - |v(\lambda, \xi) t|).
\eeq
We now resolve the step function and approximate $\rho_t(\lambda, \xi) \simeq e^{-\pi\lambda/2 + 2\xi} \Theta(\lambda - 4 \xi/\pi)$, as in Sec.~\ref{ghdql}, to rewrite this expression in terms of the double integral
\beq
C_0(t) \sim \int_{\frac{1}{2} \log t}^\infty d\xi \int_{4\xi/\pi}^\infty d \lambda e^{-\pi \lambda/2} e^{(2 - \phi) \xi} \sim t^{-\phi/2}.
\eeq
Higher energy quasiparticles give rise to a ballistic decay $1/t$ that is subleading when $\phi<2$. 
Thus, throughout the quasi-localized phase the autocorrelation function decays slower than one would expect for a model with ballistic transport. Numerical simulations of the autocorrelation function gives results in very good agreement with this exponent (Fig.~\ref{return}). We emphasize that in the argument above, it was crucial to disorder-average the full autocorrelation function---separately averaging the velocity and the density of states would yield an incorrect exponent $\phi/(2 - \phi)$ in clear disagreement with our numerical results. The local velocity is inversely proportional to the local density of states, and capturing these correlations is essential to deriving the correct anomalous exponent. 

We will show elsewhere that anomalous decay of local autocorrelation functions occurs in other models such as XXZ as well; however, the possibility of \emph{subdiffusive} behavior is specific to the disordered noninteracting models, as generic interacting models will have a finite diffusion constant due to interactions. 

\section{Discussion}

\label{sec:Discussion}

We studied the non-equilibrium dynamics of integrable spin chains with correlated disorder that preserves integrability. Focusing on the non-interacting case, we formulated a (generalized) hydrodynamic theory for such random systems, and describe the emergence of diffusive corrections due to quasiparticles scattering off random impurities. This provides a mechanism for diffusion that is different from the recent theories of diffusive corrections to GHD in {\em clean} integrable quantum systems. The predictions from hydrodynamics were compared to numerical results obtained from exact diagonalization of the free fermion problem. Both spin and energy transport can be described very accurately using hydrodynamics, provided diffusive corrections are included. Moreover, we have shown that low-energy quasiparticles are very sensitive to the tails of the disorder distribution, and can become quasi-localized, leading to an anomalous decay of local autocorrelation functions.

We expect our results to generalize naturally to all {\em interacting} random integrable systems~\cite{vega_new_1992,de_vega_thermodynamics_1994,essler_integrable_2018}. In general, we expect a complicated interplay between diffusive corrections due to disorder, and due to thermal fluctuations and interactions. However, a simpler intermediate setup would be to consider initial states for which thermal fluctuations vanish, such as the spin domain-wall initial state considered above. For such initial states, we expect our predictions to extend  naturally to the interacting case, and it would be interesting to compare the hydrodynamic predictions to matrix product state simulations.

Our results also indicate that diffusive broadening of the operator front can occur even in some non-interacting systems, which are clearly non-chaotic. These models could be used as a testbed for future diagnostic tools to distinguish chaotic from integrable systems. These models are also natural from the point of view of integrability breaking: adding integrability-breaking perturbations to a random integrable chain at strong disorder could lead to either thermalization or to many-body localization. The results of Ref.~\onlinecite{essler_integrable_2018} suggest that the regime of ballistic transport escaping localization might not be as fine-tuned as one could have expected, and it would be interesting to investigate whether hydrodynamics can still describe accurately transport away from the integrable limit.

\emph{Acknowledgments}.--- We thank Vir Bulchandani and Fabian Essler for useful comments on this manuscript. This work was supported by the US Department of Energy, Office of Science, Basic Energy Sciences, under Award No. DE-SC0019168 (U.A. and R.V.), and by NSF Grant No. DMR-1653271 (S.G.). Most numerical simulations were performed at the Massachusetts Green High Performance Computing Center (MGHPCC).

\bibliography{References}

\end{document}